\documentstyle[aps,epsf,floats,times]{revtex}
\begin{document}
\draft

\wideabs{
\title{
Human gravity-gradient noise in interferometric gravitational-wave detectors
}

\author{Kip S.\ Thorne} 

\address{
Theoretical Astrophysics, California Institute of Technology,
Pasadena, CA 91125, and}
\address{
Max-Planck-Institut f\"ur GravitationsPhysik, Schlatzweg 1,
14473 Potsdam, Germany}

\author{Carolee J.\ Winstein}

\address{Department of Biokinesiology and Physical Therapy, University of
Southern California, Los Angeles, CA 90033}
\date{Received 5 October 1998}
\maketitle
\begin{abstract}
Among all forms of routine human activity, the one which produces the strongest
gravity-gradient noise in interferometric gravitational-wave detectors
(e.g.\ LIGO) is the beginning and end of weight transfer from one
foot to the other during
walking.  The beginning and end of weight transfer entail sharp changes
(timescale $\tau \sim 20 \hbox{msec}$) in the horizontal jerk
(first time derivative of acceleration) of a person's center of mass.
These jerk pairs, occuring about twice per second, will
produce gravity-gradient noise in LIGO in the frequency band 
$2.5
\hbox{Hz} \alt
f \alt 1/(2\tau) \simeq 25 \hbox{Hz}$ with the form $\sqrt{S_h(f)} \sim
0.6 \times 10^{-23}\hbox{Hz}^{-1/2} (f/10\hbox{Hz})^{-6} \left(
\sum_i (r_i/10{\rm m})^{-6}\right)^{1/2}$.
Here the sum is over all the walking people, $r_i$ is the
distance of the $i$'th person from the nearest interferometer test mass,
and we estimate this formula to be accurate to within a factor 3.
To ensure that this noise is neglible in advanced LIGO interferometers,
people should be prevented from coming nearer to the test masses
than $r\simeq 10$m.  A $r\simeq 10$m exclusion zone will also reduce to
an acceptable level gravity gradient noise from the slamming of a door and
the striking of a fist against a wall.
The dominant gravity-gradient noise from automobiles and other vehicles
is probably that from
decelerating to rest.  To keep this below the
sensitivity of advanced LIGO interferometers will require keeping vehicles
at least 30 meters from all test masses.
\end{abstract}
}

\narrowtext

\section{Introduction and Summary}
\label{sec:Introduction}

Time-changing Newtonian gravitational forces, acting on the test masses of an
interferometric gravitational-wave detector (e.g. in LIGO), produce noise.  
This noise is
conventionally called ``gravity-gradient noise'' because the interferometer
measures the differences in the gravitational forces acting on the test
masses.  In a previous paper, Thorne and Hughes \cite{hughes_thorne} gave a 
general overview of gravity-gradient noise 
and analyzed in detail the gravity-gradient noise in LIGO due
to density fluctuations in the earth caused by
ambient seismic waves.  In this paper we focus on gravity-gradient noise due to
human activity. 

It has long been recognized that gravitational forces from moving humans 
can produce
significant noise in high-precision gravitational experiments.  
Roll, Krotkov and Dicke 
\cite{dicke} took great care to eliminate such forces in their
classic E\"otv\"os experiment, 
and Dicke \cite{dicke,leshouches}
has raised the possibility that such noise was
a serious factor in Baron Roland von E\"otv\"os's original versions of that 
experiment
\cite{eotvos}. 
In the early years of the LIGO Project, Robert Spero and others made 
rough estimates of the magnitude of human 
(and other animal) gravity-gradient noise in LIGO and the distances to 
which humans (and other animals) should be relegated to control it. 
While these estimates were sufficiently
accurate for their purposes, no analyses until ours seem to have 
identified the form of routine 
human activity that will dominate the noise (human walking), nor the spectrum 
of this dominant noise, $\sqrt{S_h} \propto f^{-6}$ for $2.5{\rm
Hz} \alt f \alt 25 {\rm Hz}$.  

A first version of our analysis was carried out in summer 1995 \cite{kip95}, 
when the
buildings that house LIGO's test masses were being designed.
Our goal was to make sure that the building design would keep humans
sufficiently far from the test masses, during routine LIGO observations, 
for human gravitational noise to be unimportant.  The press of other research
delayed until now our finalizing and publishing this analysis.  

In our 1995 document \cite{kip95}, we focused on the gravitational 
effects of a walking person's horizontal center-of-mass motion,  
and we based our analysis on two force-plate experiments taken from
the Biokinesiology (motion-in-biological-systems)
literature; cf.\ Sec.\ \ref{sec:CofMMotion} below.  
For the present paper, we have augmented our center-of-mass-motion
data base with new force-plate measurements on three
different persons; 
and using data from the Biokinesiology literature, we have verified 
our original
guess that the motion of a walking person's limbs 
produces gravitational noise small compared to that from center-of-mass
motion (Sec.\ \ref{sec:LimbMotion}).  This extended analysis has not 
changed significantly any of our 1995 conclusions.

Our analysis (Sec.\ \ref{sec:HumanWalking}) produces the following estimate
for gravity-gradient noise in LIGO due to walking people: 
\begin{eqnarray}
\sqrt{S_h(f)} &\simeq& {0.6 \times 10^{-23}\over\sqrt{\hbox{Hz}}} 
\left({10\hbox{Hz}\over f}\right)^6 
\left[ \sum_i \left({10{\rm m}\over r_i}\right)^6\right]^{1/2} 
\nonumber\\
&&\hbox{at }2.5\hbox{Hz} \alt f \alt
25\hbox{Hz}\;.
\label{eq:ShCofM}
\end{eqnarray}
Here $S_h(f)$ 
is the spectral density of the interferometer's output
gravitational-wave signal $h \ =  \Delta L / L$ (with $L = 4 \hbox{km}$ 
the interferometer arm length and $\Delta L$ the difference in arm
lengths, which fluctuates due to the gravitational forces from the walking
people); also 
$r_i$ is the distance from person $i$ to the nearest interferometer
test mass.  

We believe this estimate to be accurate to within a factor 3.  A factor 
$\sim 2$
uncertainty arises from the angular location and
direction of motion of each person [the factor $\alpha$ in Eq.\ 
(\ref{eq:alpha}) below],
and another factor that on occasion may be as large as $\sim 2$ arises from 
the gravitational forces of 
compressional and shear waves in the floor and ground, produced by the person's
walking (Sec.\ \ref{sec:FloorMotion}).  Somewhat smaller than this
are uncertainties due to variations
in the gait (walking) pattern from one person to another, and for
each person, 
from one step to another (Sec.\ \ref{sec:ForcePlateMeasurements}).  Adding
our two factor $\sim 2$ uncertainties in quadrature, we get our net factor 
$\sim 3$ uncertainty.

The $1/r_i^3$ dependence of the noise (\ref{eq:ShCofM}) 
results from the fact that it is produced by changes in the distance 
$r_i$ to the person's center of mass, and thus by the person's 
changing dipole gravitational field.  The $1/f^6$ dependence results from 
two facts: (i) The gravitational force produces a test-mass 
acceleration, which means a second time derivative of
$\Delta L$ and thence a second time derivative of $h$, and thence a $1/f^2$ 
in the amplitude spectrum $\sqrt{S_h}$. (ii) 
The fourth time derivative of the center-of-mass position is the lowest-order 
derivative that has a delta-function-like behavior on timescales
short enough to produce noise at frequences $f \sim 10$ Hz; 
and those four time derivatives produce an additional $1/f^4$ in the
spectrum.

\begin{figure}
\epsfxsize=3.3in\epsfbox{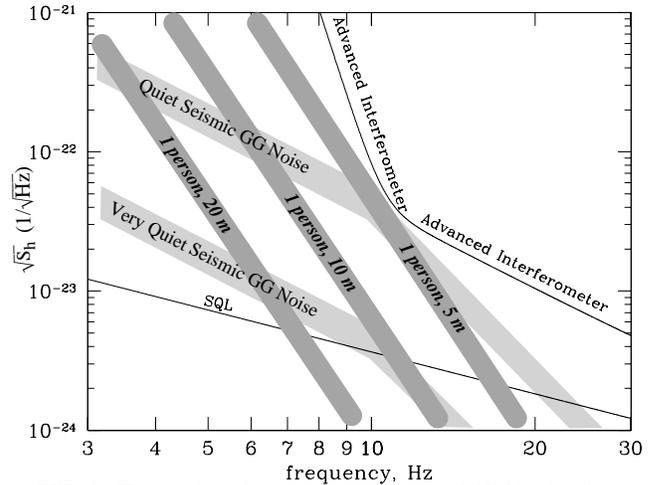}
\caption{The predicted spectrum of noise in LIGO's 4km-long interferometers
due to human gravity gradients (dark strips) and seismic gravity gradients
(light strips), compared with the benchmark noise curve for advanced LIGO
interferometers and with the standard quantum limit (SQL) in interferometers
with 1 tonne test masses.  The thickness of the strips indicates the estimated
uncertainties in our analysis.
}
\label{fig:SpectrumCM}
\end{figure}

Figure \ref{fig:SpectrumCM} shows the noise (\ref{eq:ShCofM}) 
for a single
person at various distances from the nearest test mass.  For comparison we also
show (i) the benchmark noise curve for a broad-band ``advanced'' 
interferometer (which might
operate in LIGO in the $\sim 2010$ time frame)
\cite{abramovici}, (ii) the standard quantum limit (SQL) for a LIGO
interferometer with 1 tonne test masses, and (iii) 
the estimated seismic gravity-gradient noise from ambient
earth motions \cite{hughes_thorne}
at ``quiet times'' (assuming the ``standard LIGO seismic
spectrum'') and at ``very quiet times'' (assuming a seismic spectrum
10 times lower than the standard one -- a level that might occur 
during wind-free nights). 

Figure \ref{fig:SpectrumCM} shows that a single person walking at a distance of
5m from a test mass could significantly increase the noise in an advanced
interferometer, and several people would be correspondingly more serious.

The LIGO corner building (the only one with extensive human activity) has
been designed to keep people at least 10 meters from
all test masses during normal operations.  This provides an
adequate safety factor for advanced interferometers; 
if the noise is three times as large as our estimate, then
10 people at 10 meters distance would increase an advanced interferometer's
noise by only a few tens of percent near the most sensitive frequency,
$10 \hbox{\rm Hz}$.  

It is conceivable that in $\sim 2010$ or later interferometers will be
operated in LIGO with good performance at frequencies below 10 Hz---e.g., as
low as $\sim 3$ Hz.  (Adalberto Giazzotto and colleagues of the VIRGO Project
have developed seismic isolation systems that can go down to frequencies as low
as 3 Hz \cite{giazzotto}.)  Figure \ref{fig:SpectrumCM} indicates that, if such
interferometers are ever operated in LIGO, it will be necessary to expand the
people-free zone around each corner test mass.

The LIGO end and mid-station buildings (which have little human activity)
are designed to keep all humans at least 5 meters from the end test masses
during normal operations.  This provides an adequate safety margin for the
first and enhanced LIGO interferometers, which will operate in the early 
and mid 2000's; but when advanced interferometers begin to operate, 
it will be necessary to expand the people-free zone, most especially at 
the end of each end building.  

Robert Spero's early estimates of human gravity gradient noise focused on 
the slamming of a door or the striking of a fist against a wall \cite{spero}. 
Since this is more readily suppressed (by warnings and viscous door stops)
than human walking, we have regarded it as a less serious and pervasive noise
source.  However, whenever a door slams or a fist strikes, the magnitude of 
the resulting gravitational ``signal'' in an 
advanced LIGO interferometer will be comparable to that from people walking:

In Sec.\ \ref{sec:AutomobileMotion}, 
by a variant of Spero's analysis we derive the following 
expression for the Fourier
transform of the signal $h(t)$ produced by a mass $M$ 
striking a building wall and coming suddenly to rest:
\begin{equation}
| \tilde h | = {G M |\alpha \Delta v | \over L r^3 (2\pi f)^4} \;.
\label{eq:tildehDoorFist}
\end{equation}
Here $\Delta v$ is the object's sudden change of speed, $r$ is its distance
from the nearest interferometer test mass, $L=4$km is the interferometer
arm length, 
and $\alpha$ is a coefficient in the range $-2 \alt \alpha \alt +2$ that
depends on object's angular location.  Following Spero, 
we show that, 
with optimal signal processing, this ``signal'' would produce the following
amplitude signal-to-noise ratio in an advanced LIGO interferometer:
\begin{equation}
{S\over N} \simeq 1 \left({M \Delta v\over5 \hbox{kg m/s}}\right)
\left(10{\rm m}\over r\right)^3\;.
\label{eq:SOverNFistDoor}
\end{equation}
Here our fiducial value, 5 kg m/s, for $M \Delta v$ corresponds to a 5 kg door
slamming shut at a speed of 1 m/s, or a 2 kg fist and forearm striking a 
wall at a speed of 2.5 m/s.  

Thus, slamming doors and striking fists, like walking people, must be kept at
distances $r \agt 10$m from the test masses of advanced LIGO interferometers. 

A third type of human activity that can produce strong gravitational noise is
the motion of automobiles and other vehicles.  In Sec.\ 
\ref{sec:AutomobileMotion} we argue that the dominant vehicle
gravity-gradient noise, in the critical frequency region $f\sim 10$ Hz, is due
to a vehicle's sudden change of acceleration when it comes to rest, e.g.\ when
parking.  If $\Delta a$ is the vehicle's change of acceleration, 
$M$ is its mass, and $r$ is its distance from the nearest interferometer
test mass, 
then in an advanced LIGO
interferometer the vehicle 
will produce a ``signal'' $h(t)$ with Fourier transform 
in our frequency band
\begin{equation}
| \tilde h | = {GM |\alpha\Delta a| \over L r^3 (2\pi f)^5} \;.
\label{eq:tildehAutomobile}
\end{equation}
Here $\alpha$ is the same angle-dependent
coefficient with range $-2 \alt \alpha \alt +2$ as appears in the door/fist
analysis.  With optimal signal processing, this
``signal'' would produce the following amplitude
signal-to-noise ratio in an advanced LIGO interferometer:
\begin{equation}
{S\over N} \simeq 1 \left({M\over2 \hbox{tonne}}\right) 
\left({|\Delta a|\over 0.6 g}\right)  
\left(30{\rm m}\over r\right)^3\;.  
\label{eq:SOverNAutomobile}
\end{equation}
Here $g = 9.8 {\rm m} {\rm s}^{-2}$ is the acceleration of gravity and
we have set $|\alpha|$ to a representative value, $\sqrt2$.

The LIGO service roads come no closer than 40 meters 
to a LIGO corner test mass,
but they approach to within 15 meters of the corner- and mid-station test
masses. 

The product $M |\Delta a| = 2 {\rm tonne} \times 0.6 g$ used in Eq.\
(\ref{eq:SOverNAutomobile}) is in the upper range of what one might 
expect for
stopping vehicles.  Two tonnes is a modest vehicle mass; $0.6g$ is the
deceleration at which a vehicle begins to skid on dry asphalt.  Thus, when
advanced interferometers are operating in LIGO it will be necessary to increase
the radius of the vehicle-free zone at the corner and mid stations to $\agt
30$m.  

The body of this paper is organized as follows:  At the beginning of
Sec.\ \ref{sec:HumanWalking}
we briefly explain why human walking is the dominant source of human gravity
gradient noise in interferometers, and why, at the frequencies of interest,
the noise comes predominantly from sudden changes in motion. 
Then in Sec.\  \ref{sec:CofMMotion} we
compute the gravity-gradient noise in an interferometer due to a person's
center-of-mass motion, and sum over a population of people to get Eq.\
(\ref{eq:ShCofM}) (discussed above).  Momentum conservation implies that any
sudden change in a person's center-of-mass motion will produce a corresponding
sudden displacement of the floor and the ground beneath the floor; in 
Sec.\  \ref{sec:FloorMotion} and an Appendix
we show that gravity gradient noise from
this floor/ground motion will {\it not} cancel that from the person, but 
{\it can}, on occasion, cancel as much as half of it
(thereby introducing a factor $\sim 2$ uncertainty in the net noise). 
In Sec.\  \ref{sec:LimbMotion} we
compute the gravity-gradient noise due to the motions of a person's
limbs (including, most importantly, the sudden changes of motion
when a heel strikes the floor).  We show that the limb motion produces noise 
that is smaller by a factor $\sim 0.1 (10{\rm m} /r)(f/10{\rm Hz})$
than the noise from center-of-mass motion; here $r$ is the distance 
between the person
and the nearest test mass.  In Sec.\ \ref{sec:AutomobileMotion} we analyze the
gravity-gradient noise due to the sudden stopping of a moving mass---a
slamming door, a fist striking a wall, or a parking vehicle---arriving
at Eqs.\ 
(\ref{eq:SOverNFistDoor}) and (\ref{eq:SOverNAutomobile}), discussed above.
In Sec.\ \ref{sec:Conclusions} we make some concluding remarks.

\section{Human Walking}
\label{sec:HumanWalking}

Consider a person (or vehicle) moving in the vicinity of a LIGO test mass.
Denote by $\Phi$ the person's Newtonian gravitational potential in the
test-mass vicinity, and by $\vec x' (t)$ and $r'(t) = |\vec x' |$ 
the vector and distance from the person's center of
mass (``CofM'')
to the test mass at time $t$.  Expand the Newtonian potential in 
multipole moments around the person's (moving) CofM:
\begin{equation}
\Phi = - {GM\over r'} - {3\over 2} {G {\cal I}_{jk}\over r'^5} x'_j x'_k +
\ldots\;.
\label{eq:PhiTotal}
\end{equation}
Here ${\cal I}_{jk}(t)$, the person's 
quadrupole moment relative to his CofM, is the means by which
his moving limbs produce gravity-gradient noise, repeated indices ($j$ and $k$)
are to be summed, and we use Cartesian coordinates so it doesn't matter whether
tensor indices are up or down. 

In Sec.\ \ref{sec:CofMMotion} we examine the first (monopolar) term in Eq.\
(\ref{eq:PhiTotal}).  This is the dominant gravitational effect of the
CofM motion.  In Sec.\ \ref{sec:LimbMotion} we examine the second
(quadrupolar) term---including its time dependence due both to limb motions 
and to 
center-of-mass motions---and we show that its influence on an interferometer's
noise is small compared to that of the monopolar term.

Before presenting these analyses, it may be useful to comment on the 
importance of sudden changes (jerkiness) in the human (or vehicular) motion. 

Smooth (non-jerky) motion produces gravitational forces 
that are concentrated at
frequencies $f \sim 1$ Hz, well below those, $f \agt 10$ Hz, of concern for the
interferometers.  For example, the period of the normal
human gait cycle (two steps,
one left and one right) is about 1 sec (frequency 1 Hz); and an autumobile 
moving at speed 30km/hr at a distance 15m from a test mass travels through an
angle $90^{\rm o}$ as seen by the test mass in about 2 sec (frequency $\sim
0.5$ Hz).  If the motion is sufficiently smooth, the Fourier transform of 
such motions will fall off with frequency exponentially, 
becoming totally negligible at $f\sim 10$ Hz.  

By contrast, if
the $n$'th time derivative of the motion changes significantly on a timescale
$\tau \alt 50$msec, then the $n+1$'th time derivative will have a sharp,
delta-function-like peak with time width $\tau$, and that will produce a
Fourier transform of the motion that falls off as $1/f^{n+1}$ up to frequencies
$f \sim 0.5/\tau \agt 10 {\rm Hz}$.  Such a power-law fall-off can produce 
much larger gravity-gradient noise near 10 Hz than the  exponential fall-off
caused by smooth motion.

\subsection{Motion of Center of Mass}
\label{sec:CofMMotion}

In this section we examine the gravitational noise produced by the monopolar
(center-of-mass) term in Eq.\ (\ref{eq:PhiTotal}). 

\subsubsection{General formulas for noise}
\label{sec:GeneralFormulas}

We focus on
a time duration of one gait cycle $\simeq 1$sec, centered on time $t=0$.  
We introduce Cartesian coordinates $x_j$ attached to the floor, with origin 
at the location of the person's CofM at time $t=0$; and we denote by
$\vec \xi (t)$ the motion of the CofM relative to this origin (so
$\vec\xi=0$ at $t=0$).  Then the vector and distance $\vec x'$ and $r'$
from the person's CofM to the test mass are
\begin{eqnarray}
\vec x' &=& \vec x - \vec \xi\;, \nonumber\\
r' &=& |\vec x - \vec \xi| = r - \hat n
\cdot \vec \xi + {\vec\xi^2 - (\hat n \cdot \vec \xi)^2 \over 2r} + \ldots\;.
\label{eq:vecxprime}
\end{eqnarray}
Here $\vec x$ and $r$ are the vector and distance from the origin of 
coordinates to the test mass, and 
$\hat n = \vec x/r$ is the unit vector pointing from the origin to 
the test mass.
Correspondingly, we can write the center-of-mass piece of the gravitational
potential (\ref{eq:PhiTotal}) as
\begin{eqnarray}
\Phi_{\rm CofM} &\equiv& -{GM\over r'} = - {GM\over r} 
- {GM \xi_j n_j \over r^2} 
\nonumber\\
&&\quad\quad
- {3\over 2} {GM(\xi_j \xi_k - {1\over 3} |\vec\xi|^2 \delta_{jk})n_jn_k\over
r^3} + \ldots \;. 
\label{eq:PhiCM}
\end{eqnarray}

The first term (monopolar about our fixed center of coordinates)
is constant in time and thus can produce no gravity-gradient noise, so we shall
ignore it.  The second
term (dipolar about our fixed center of coordinates) produces the 
gravity-gradient noise via sudden
changes of the CofM location $\vec\xi$. 
The third term (quadrupolar about our fixed center of coordinates) is smaller
than the second, dipolar term by $\sim |\vec\xi|/r \alt 0.7 {\rm m}/10{\rm m}
\sim 0.1$, and its gravity-gradient noise is correspondingly smaller in our
frequency band, so we shall ignore it.
The gravitational acceleration of the test mass, produced by the person's
CofM motion, is then minus the gradient of the second, dipolar term:  
\begin{equation}
g_j = {GM\over r^3} ( \xi_j - 3 n_j n_k \xi_k)\;.
\label{eq:gj}
\end{equation}

\begin{figure}
\epsfxsize=2in\epsfbox{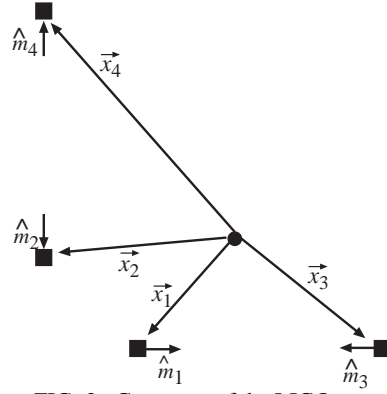}
\caption{Geometry of the LIGO test masses (solid squares)
and the location (large dot) of the
center of mass of a person at time $t=0$.
}
\label{fig:interferometer}
\end{figure}

The interferometer has four test masses labeled $A=1,2,3,4$, each of which 
experiences an acceleration of the form (\ref{eq:gj}) due to the CofM
motion.  The resulting output signal of the interferometer, $h(t) = \Delta
L(t)/L$, has a second time derivative given by 
the sum of these accelerations projected
onto unit vectors $\hat m_A$
that point along the interferometer arms as shown in Fig.\
\ref{fig:interferometer}:
\begin{eqnarray}
{d^2 h \over dt^2} &=&  {1\over L} \sum_A \vec g_A \cdot \hat m_A  
\nonumber\\
&=& \sum_A {GM\over Lr_A^3} 
\left[ \vec\xi \cdot\hat m_A  
- 3 (\hat n_A \cdot \hat m_A)(\hat n_A \cdot \vec\xi)\right] \;.
\label{eq:dhdt}
\end{eqnarray}

Since the interferometer's laser beams are horizontal, the vectors $\hat m_A$
are all horizontal; and since the person's CofM is at the same
height as the test masses to within $< 1$ meter and the
person is at a distance $\sim 10$ m from the nearest test mass, the vertical
component of $\hat n_A$ is $\alt 0.1$ of the horizontal component.  
This means [cf.\ Eq.\ (\ref{eq:dhdt})] that the vertical component
of $\vec\xi$ produces significantly less gravitational
noise than the horizontal component: its contribution to $\sqrt{S_h}$ 
is less by a factor $\alt 0.1 |\tilde \xi_{\rm v}|/|\tilde \xi_{\rm h}| 
\sim 0.3$.  (Here the tilde denotes a
Fourier transform, v and h denote vertical and horizontal components,
and we have used
$|\tilde \xi_{\rm v}|/|\tilde \xi_{\rm h}| 
\sim 3$ at $f\sim 10$ Hz as inferred
from force-plate measurements; Table \ref{table:Force} below.)  On this basis,
we shall ignore the vertical component of the CofM motion
and approximate $\vec\xi$ and $\hat n_A$ as purely horizontal. 

It will be convenient below to rewrite Eq.\ (\ref{eq:dhdt}) in the form
\begin{equation}
{d^2h\over dt^2} = \alpha {GM  \xi \over L r^3}\;,
\label{eq:dhdt1}
\end{equation}
where $r$ is the distance from the center of coordinates to the nearest 
test mass, $\xi (t)$ (scalar, not vector) is the distance the CofM 
has traveled since $t=0$
($\xi = -|\vec\xi|$ for negative times and $+|\vec\xi|$ for positive), 
and $\alpha$ is a dimensionless
coefficient given by
\begin{equation}
\alpha = 
\sum_A \left({r\over r_A}\right)^3 \left[ \hat\xi \cdot\hat m_A  
- 3 (\hat n_A \cdot \hat m_A)(\hat n_A \cdot \hat\xi)\right] \;.
\label{eq:alpha}
\end{equation}
Here $\hat\xi = \vec\xi / \xi$ is the unit vector along the CofM's
direction of motion, which we regard as constant during one gait cycle. 

For people near an end mass (mass $A=3$ or 4 in Fig.\
\ref{fig:interferometer}), only that mass gives a significant contribution to
the coefficient $\alpha$, 
and it is easy to verify that $\alpha$ ranges from -2 to +2,
depending on the angular location of the person.  The extremal values $\pm 2$
are reached when the person's CofM is along the interferometer arm
and the person is moving toward or away from the test mass ($\vec \xi$, $\hat
m_A$, and $\hat n_A$ all parallel or antiparallel).

For people in LIGO's corner station, both corner masses contribute strongly to
$\alpha$.  This is because
the distance between the two corner masses, $l\simeq 5$m, is comparable
to the distance between the person and the nearest test mass, 
$r \sim 10$m. A straightforward numerical exploration shows, in this case, that
$\alpha$ ranges from about -2.2 to about +2.2 depending on the 
person's location.  

A representative value for $|\alpha|$, which we use in our final noise
estimates [e.g., Eq.\ (\ref{eq:ShCofM}) above], is 
$|\alpha|_{\rm representative}
= \sqrt2$.

\begin{figure}
\epsfxsize=3.3in\epsfbox{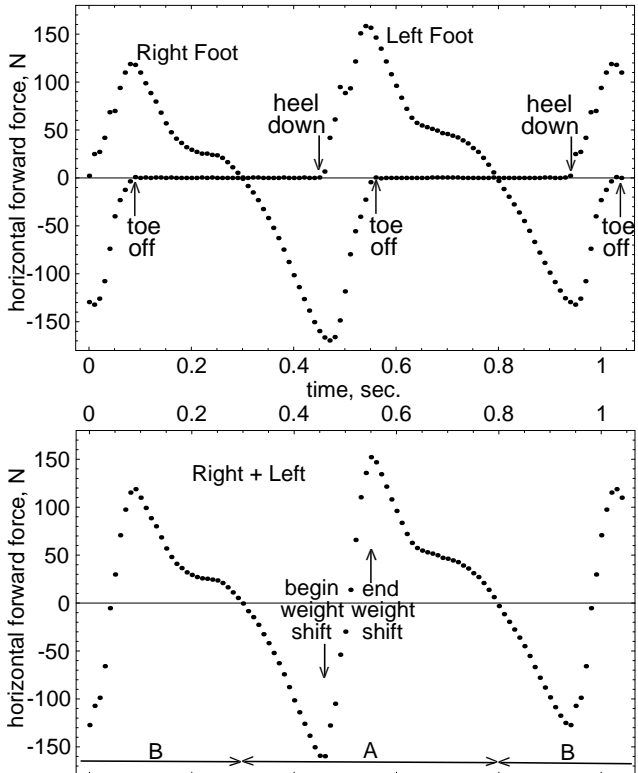}
\caption{The forward, horizontal force exerted on the floor by a woman 
weighing 73 kg, as measured using dual (two-feet) force plates by
Earnest L. Bontrager \protect\cite{bontrager} (data set 3506a5).  
Upper panel: The force as a function
of time exerted by each foot.  Lower panel: The sum of the forces from the two
feet; the full gait cycle is divided into two half cycles, A and B. 
}
\label{fig:Force} 
\end{figure}

Eq.\ (\ref{eq:dhdt1}) implies that the noise spectrum $\sqrt{S_h(f)}$ will scale
with frequency as $|\tilde\xi(f)|/f^2$, where $\tilde\xi$ 
is the Fourier transform
of the distance traveled $\xi(t)$; and the discussion just before the beginning
of this section implies that $\tilde\xi(f)$ will be governed primarily by the
lowest order derivative of $\xi(t)$ that has sudden changes on a timescale
$\tau \alt 50$ msec.  To identify that derivative and the details of the sudden
changes, we rely on experimental data from the field of Biokinesiology (the
study of motion in biological systems).

Since $\xi(t)$ is distance traveled by the CofM, the only way that it
or its derivatives can change jerkily is by the application of a sharply
changing, horizontal external force.  The only such force, as a person walks,
is the horizontal force of the floor on the person's feet.  The negative of
that force (the horizontal force $F(t)$ of the feet on the floor) is measured
by biokinesiologists, using force plates (pp.\ 414--418 of Ref.\ \cite{perry};
Sec.\ 4.2 of Ref.\ \cite{winter}).  By
momentum conservation, this measured force is $F = - M d^2\xi/dt^2$, and
correspondingly the gravitational noise is related to the measured horizontal
force by
\begin{equation}
{d^4h\over dt^4} = \alpha {G F\over L r^3}\;.
\label{eq:d4hdt4}
\end{equation}

\subsubsection{Force-plate measurements}
\label{sec:ForcePlateMeasurements}

Figure \ref{fig:Force} 
shows the measured horizontal force $F(t)$ exerted on the floor
by a woman weighing 73 kg during a full gait cycle.  These data were obtained 
as follows: 

Our colleague, Earnest L.\ Bontrager placed two force plates
in the floor of his laboratory at Rancho Los
Amigos Medical Center, Downey, California.  The force plates were so located
that in normal walking a person will encounter them during one gait cycle,
with the right foot landing on the first plate and then the left foot 
on the second.
Each plate was equipped with piezzo-electric transducers that measured the 
foot's forward horizontal (``progressive'') force, its vertical force, and 
its sideward horizontal (``medial'') force.  Figure \ref{fig:Force} is based on
the progressive force measurements.  The transducer outputs were sampled and
recorded at 2500 samples per second and were then averaged over 0.01 second
intervals to produce, for each measured half-gait cycle, a single data set. 
Figure \ref{fig:Force} is based on Bontrager's data set 3506a5
(data set a5 for person 3506) \cite{bontrager}.

For Fig.\ \ref{fig:Force} we modified the data set by adding, at the beginning,
the last 9 points from the left-foot measurement; and at the end, the first 11
points from the right-foot measurement (under the plausible assumption that the
unmeasured end of the previous left-foot gait cycle is the same as the measured
cycle, and the unmeasured beginning of the next right-foot cycle is the same as
the measured cycle).  We have divided the full gait cycle into two half 
cycles A and B, as shown in Fig.\ \ref{fig:Force}.

Equation (\ref{eq:d4hdt4}) implies that the noise spectrum $S_h(f)$ is
proportional to $|\tilde F(f)|/f^4$, where $\tilde F$ is the Fourier transform 
of $F(t)$, the sum of the
forces from the two feet (bottom panel of Fig.\ \ref{fig:Force}).  As was
discussed at the beginning of Sec.\ \ref{sec:HumanWalking}, the only
features of $F(t)$ that can contribute significantly at frequencies $f\sim
10$Hz are those that change on timescales $\tau \alt 50$ msec.  The net force
$F(t)$ varies only modestly on such short timescales, but its first time 
derivative $dF/dt \equiv \dot F$ varies strongly: Shortly after placing his
heel on the floor (``heel down''), the
person begins to transfer weight from his trailing foot to his leading foot; 
this beginning of weight transfer
entails a change $\Delta \dot F \simeq 4000 {\rm N/s}$ 
in the slope of the force curve (change of jerk) on a very short timescale
$\tau \sim 20$ msec.  A time $\delta t \simeq 90$ msec later,
just before the trailing toe lifts off the ground (``toe off''), 
the weight transfer 
ends with
a second sharp change of jerk $-\Delta\dot F$.  

These two jerk changes during each half gait cycle
give rise to the following (approximate) form of the half-cycle's 
Fourier transform $\tilde F_{\rm 1/2}$ in the
frequency band of interest, $2.5 {\rm Hz} \alt f \alt 25 {\rm Hz} \simeq
0.5/\tau$:
\begin{equation}
|\tilde F_{1/2}| = 
{1\over (2 \pi f)^2} 2 \sin(\pi f \delta t) \Delta\dot F \;.
\nonumber
\end{equation}
A numerical Fourier transform (FFT) of $F(t)$ (Fig.\ \ref{fig:Force})
reveals that other features
with timescales $\agt \delta t = 90$ msec produce modulations of $\tilde F(f)$
analogous to but no larger than the $\sin(\pi f \delta t)$.  We shall ignore
these modulations and correspondingly shall approximate $\sin(\pi f \delta t)$
by its rms value, $1/\sqrt2$, and therefore shall rewrite the above formula as
\begin{equation}
|\tilde F_{1/2}| = {\sqrt2 \Delta\dot F \over (2\pi f)^2}\;.
\label{eq:tildeF}
\end{equation}

Equation (\ref{eq:tildeF}) is a fairly accurate representation of 
the half-gait-cycle Fourier transform, not just for the data set in Fig.\
\ref{fig:Force}, but for all force-plate data that we have examined --- data in
the Biokinesiology literature \cite{perry,winter}, and unpublished data on 
two other subjects in Bontrager's laboratory \cite{bontrager}.  

In all of Bontrager's data sets except 3506a5 (Fig.\ \ref{fig:Force}), 
only one force plate was used rather than two, so the data sets show 
only the horizontal force produced on the floor by one foot during a half gait 
cycle, along with the times of heel down and toe off for the second foot as
measured by a switch attached to the foot. To compute the force of the second
foot, we assumed (in accord with Bontrager's advice) that its force
history was the same as that measured for the first, but displaced in time
as shown on the foot-switch recordings.  We thereby modified each 
data set to include the force of the second foot, and computed
the total force of the two feet (analog of segment A and segment B of
Fig.\ \ref{fig:Force}).

We have fit Eq.\ (\ref{eq:tildeF}) to the Fourier transforms of
each of Bontrager's 
data sets for the total force of both feet during a half gait cycle.  For
each single-force-plate data set, and for the two half-gait-cycles (A and B)
of dual-plate set 3506a5, we did a least-squares solution for
$\Delta \dot F = |\tilde F_{1/2}|(2\pi f)^2 / \sqrt2$, and 
its rms fluctuations in the frequency range 2.5 to 20Hz.  The results are 
shown in Table \ref{table:Force}.  

\begin{table}
\caption{Sudden changes of jerk, and ratio of vertical to horizontal force
spectra for a half gait cycle, as inferred from Bontrager's data sets 
\protect\cite{bontrager}.  
}
\label{table:Force}
\vskip15pt
\begin{tabular}{llll}
Subject&Plates&$\Delta {dF/dt}$&${|\tilde 
F_{\rm v}|/|\tilde F_{\rm h}|}$\\
\tableline
Male 82kg&&&\\
\hskip6pt 3700c5&single&$6400\pm1900$&1.7\\
\hskip6pt 3700c7&single&$6000\pm3400$&2.6\\
Male 78kg&&&\\
\hskip6pt 3772c5&single&$4600\pm1300$&3\\
\hskip6pt 3772c7&single&$7100\pm2600$&3\\
Female 73kg&&&\\
\hskip6pt 3506a5 --- A&dual&$4400\pm2200$&5\\
\hskip6pt 3506a5 --- B&dual&$3700\pm1900$&4\\
\end{tabular}
\end{table}

Note that in each measured spectrum there are fluctuations of
typical magnitude 30--60 per cent around the $1/f^2$ law of Eq.
(\ref{eq:tildeF}); and the fluctuations in the inferred $\Delta\dot F $ from 
one gait cycle to another and from person to person are of order 30 per cent. 
If a person were to run rather than walk, the resulting $\Delta\dot F$ might be
larger, but presumably not by more than a factor $\sim2$; and running in the
vicinity of a LIGO test mass should be much less common than walking.
These variations in $\Delta\dot F$ are modest contributors to our overall 
factor $\sim 3$ uncertainty in the gravity gradient noise.
Based on Table \ref{table:Force}, we shall use the value 
\begin{equation}
\Delta\dot F  = 5500 \hbox{N/s}
\label{eq:DeltadFdt}
\end{equation}
in our noise evaluations.

For Bontrager's data sets we have also evaluated the ratio $|\tilde F_{\rm v}|
/|\tilde F_{\rm h}|$ of the Fourier transforms of vertical force and 
horizontal force in the vicinity of 10 Hz; see the last column of Table
\ref{table:Force}.  The vertical spectra vary by a large factor from one person
to another: male 3700 walks rather smoothly; female 3506 strikes the floor 
sharply
with her heel at heel down.  Correspondingly, female 3506 produces a
large change of force $\Delta F_{\rm v}$ at heel down 
on a short enough timescale to dominate
the vertical spectrum, $\tilde F_{\rm v} 
\simeq (2\pi f)^{-1} \Delta F_{\rm v}$; whereas male 3700 (and also male 3772)
has a vertical spectrum dominated by a change of jerk and therefore falling off
more rapidly with frequency, $|\tilde F_{\rm v}| \simeq (2\pi f)^{-2} 
\Delta\dot F_{\rm v}$.  
As a result, female 3506 has a significantly larger ratio of
vertical to horizontal $|\tilde F|$ than the males.  
However, even for her large heel-induced
$|\tilde F_{\rm v}|$, the resulting vertical contribution to the gravity 
gradient
noise is smaller than the horizontal contribution --- smaller by the factor
$\sim 0.1 |\tilde F_{\rm v}|/|\tilde F_{\rm h}| \sim 0.5$ discussed in
Sec.\ \ref{sec:GeneralFormulas}. 

\subsubsection{Noise spectrum}
\label{sec:NoiseSpectrum}

For people walking in the vicinity of LIGO test masses, the sharp changes of
jerk are not likely to occur in a periodic fashion to within a period accuracy
of 0.05 sec, and correspondingly, in the vicinity of 10 Hz the jerks are not
likely to superpose coherently.  Therefore, we can approximate the sharp
changes of jerk as constituting a random shot noise, for which the spectral
density of the gravitational-wave noise from a single person
will be 
\begin{equation}
\sqrt{S_h} = \left(2 {2\over P_{\rm gait}}\right)^{1/2} 
| \tilde h_{1/2}| = 
{2\sqrt2 \alpha G \Delta \dot F \over L \sqrt{P_{\rm gait}} r^3 (2\pi f)^6}\;,
\label{eq:ShOnePerson}
\end{equation}
Here $P_{\rm gait}$ is the gait-cycle period (about 1 second). In the first
expression $2/P_{\rm gait}$ is the rate of (dual-jerk) ``shots'' (half gait
cycles) for each of which $ \tilde h_{1/2}$ is the Fourier transform of
$h(t)$, and the second expression follows from Eqs.\ (\ref{eq:d4hdt4}) and
(\ref{eq:tildeF}). 

For a number of walking people, each at a different distance $r_i$
from the test mass and with a different angular location and direction of
motion and corresponding factor $\alpha_i$, the noises will add in quadrature,
producing 
\begin{equation}
\sqrt{S_h} = {2\sqrt2 G \Delta\dot F \over L \sqrt{P_{\rm gait}}(2\pi f)^6}
\left(\sum_i {\alpha_i^2 \over r_i^6}\right)^{1/2}
\label{eq:ShPeople}
\end{equation}
Inserting the numerical values (discussed above) $|\alpha_i| = 
|\alpha|_{\rm
representative} = \sqrt2$, $\Delta \dot F = 5500 {\rm N/s}$, $L=4 {\rm km}$,
$P_{\rm gait} = 1 \rm{s}$, we obtain the noise spectrum (\ref{eq:ShCofM})
discussed in the introduction and the abstract. 
  
\subsection{Motion of Floor and Ground}
\label{sec:FloorMotion} 

Sharp changes in the horizontal force $F(t)$ 
will produce deformations of the floor and ground, which 
become seismic waves as they propagate out through the earth.  These 
deformations will produce gravity gradient noise correlated with that 
from the person's CofM.  In this section we shall estimate this noise.

We begin with the standard expression 
\cite{hughes_thorne} for the gravitational potential at test-mass location
$\vec x_A$, produced by the displacement $\vec \zeta$ of the floor and ground: 
\begin{equation}
\Phi = - \int_{\cal V} 
{-G\;\vec\nabla\cdot (\rho\vec\zeta)\over |\vec x - \vec x_A|} d^3 x 
- \int_{\cal \partial V} {G\;\rho \vec\zeta \cdot d\vec A \over |\vec x - \vec
x_A|}\;.
\label{eq:Phideltarho}
\end{equation}
Here $-\vec\nabla\cdot(\rho\vec\zeta)$ is the Eulerian change in density
induced by the displacement $\vec\zeta$, the first integral is over the
interior $\cal V$ of the floor and ground, and the second integral is over the
surface layer of mass produced by $\vec\zeta$ on the surface $\partial {\cal
V}$ of the floor and the adjoining ground outside the LIGO buildings.

Integrating the first term by parts and cancelling the resulting surface term
against the second term of (\ref{eq:Phideltarho}), we obtain
\begin{equation}
\Phi = - \int_{\cal V} G\rho \zeta_j \left( {1\over | \vec x - \vec
x_A|}\right)_{,j} d^3x\;.
\label{eq:Phidipole}
\end{equation}
The gravitational acceleration on test-mass A is minus the gradient of this
with respect to $\vec x_A$.  Since the only dependence of $\Phi$ on $\vec x_A$
is through the combination $\vec x - \vec x_A$, the gradient can be replaced by
a derivative under the integral with respect to $- \vec x$, thereby giving
\begin{equation}
g_i = - \int_{\cal V} G\rho \zeta_j \left( {1\over |\vec x - \vec x_A |}
\right)_{,ji} d^3 x\;.
\label{eq:gFloor}
\end{equation}
This is the dipolar floor/ground analog of Eq.\ (\ref{eq:gj})
for the gravitational acceleration produced by the person's CofM.  By the
procedure that led from (\ref{eq:gj}) to (\ref{eq:dhdt}), we obtain the
gravitational-wave noise $d^2h/dt^2$ produced by the floor/ground motion, 
to which we add the person's CofM noise (\ref{eq:dhdt}).  Differentiating the
result once in time, we obtain
\begin{eqnarray}
&&{d^3 h \over dt^3} = -{G\over L}  \sum_A m_{Ai} 
\nonumber\\ 
&&\quad\times \left[ M \dot 
\xi_j \left( {1\over | \vec x
- \vec x_A |} \right)_{,ji} 
 + \int_{\cal V} \rho \dot\zeta_j 
\left( {1\over | \vec x - \vec x_A |} \right)_{,ji} d^3x \right]\;.
\nonumber\\
\label{eq:d3h3Floor}
\end{eqnarray}
Here (as should be obvious) in the first term $\vec x$ is the person's CofM
location and in the second it is a location in the floor or ground.

This equation expresses the gravitational noise in terms of
sharp changes in the person's CofM momentum and the 
momentum density of the ground.  By momentum conservation, any change of
the CofM momentum $M \dot\xi_j$ must be accompanied by an equal and opposite
change of the total floor/ground momentum 
$\int_{\cal V} \rho \dot \zeta_j
d^3x$.  If the suddenly deposited momentum remains close to the person
[within a distance $\ll r = ($person's distance to nearest test mass)] during a
time $\tau \simeq 1/(2 f) \simeq 50 \rm{msec}$, then the gravity-gradient 
noise from the floor/ground will nearly
cancel that from the person's CofM.  If the
deposited momentum spreads out over a distance $\gg r$ in the time $\tau$, then
it will produce a neglible gravitational force on the test mass, and negligible
gravity gradient noise.

The deposited momentum moves outward through the floor and ground with 
seismic-wave speeds; it resides in a spreading, widening shell whose sharp
outer
edge moves at the seismic P-wave speed $c_P$ and fuzzy
inner edge at a little less
than the seismic S-wave speed $c_S$ (see the Appendix). 

The floor on which the person walks is a slab of reinforced concrete 20cm
thick.  In each corner and end station this slab begins about 6m from the test
mass and extends outward to about 18m from the test mass, and transversely
about 12m in each direction.  In the corner station the slab begins about 
10m from the nearest
test mass and extends on outward an additional 
10 to several 10's of meters, in a
complicated shape.  The concrete has $c_P \simeq
3700$m/s and $c_S \simeq 1700$m/s; so in $\tau \simeq 50 \rm{msec}$, the outer
edge of the spreading shell could move a horizontal distance of 180m and the
inner edge, 85m if the slab were that large.  To the extent, then, that the
deposited momentum gets trapped in the slab for $\tau\sim 50$msec, it spreads 
through the whole slab;
and since most of the slab is somewhat farther 
from the test mass than the nearest
person (about 10 meters) and in somewhat different directions (off to the
sides),
the slab's sudden momentum change will produce a considerably 
weaker gravitational force on
the test mass than is produced by the person's sudden momentum change.

The momentum
that passes through the thin floor and into the ground below spreads
through the ground at much lower speeds than that confined to the floor: 
$c_S \simeq
270$m/s; $c_P \simeq 520$m/s in Hanford's dry soils and 1700m/s in 
Livingston's water-saturated soils; 
cf.\ Tables II and IV of Ref.\ \cite{hughes_thorne}.  
Correspondingly, in $\tau = 50$msec time, the inner edge of the
spreading momentum shell travels a distance $\simeq 13$m; and the outer edge,
$\simeq 25$m at Hanford and $\simeq 80$m at Livingston.  These distances, being
comparable to, and much larger than the person's $\sim 10$m separation from the
test mass, will cause the momentum suddenly deposited in the ground to produce 
a somewhat smaller gravitational noise than that of the person; there is no
possibility for a strong cancellation. 

On the other hand, in some cases the Green's function for the momentum spread 
(cf.\ Appendix, and Figs. 2--4 of Ref.\ \cite{johnson}) is strongly localized
near the inner edge of the spreading shell.  In such cases, with the momentum
having spread only a distance $\sim 15$m compared to the person's $\sim 10$m
distance from the test mass, the ground's gravitational force on the 
test mass might be as much as half that of the person, thereby reducing the net
noise by a factor 2 relative to that of the person alone.  This is a
significant
contributor to our estimated factor 3 uncertainty in the net noise.

\subsection{Motion of Limbs}
\label{sec:LimbMotion}

Turn, next, to the noise produced by the person's quadrupolar gravitational
field
\begin{equation}
\Phi_{\rm Quad} = - {3\over 2} {G {\cal I}_{jk}\over r'^5} x'_j x'_k
\label{eq:PhiQuad1}
\end{equation}
[Eq.\ (\ref{eq:PhiTotal})].  Here ${\cal I}_{jk}$ is the
quadrupole moment about the person's moving CofM, and $x'_j$ and $r' = |\vec
x'|$ are the vector and distance from the CofM to a test mass.  
Replacing $x'_j$ by $x_j - \xi_j$, where $x_j$ reaches from the (fixed) CofM
location at the center of a gait cycle to the test mass,
and $\xi_j(t)$ is the displacement
of the CofM relative to that fixed point, we obtain
\begin{equation}
\Phi_{\rm Quad} = -{3\over2} {G{\cal I}_{jk} n_j n_k \over r^3} - {3\over2}
{G{\cal I}_{jk}\over r^4} (3 n_j n_k \xi^{||} - 2 n_j \xi^{\perp}_k )\;,
\label{eq:PhiQuad}
\end{equation}
where $\xi^{||}$ is the component of $\vec\xi$ along $\hat n = \vec x/r$, 
the direction to the test mass, and $\xi_k^{\perp}$ 
is its projection orthogonal
to $\hat n$.

This quadrupolar gravitational field produces interferometer noise via jerky
changes of ${\cal I}_{jk}$ (noise ``intrinsic'' to the person's quadrupole
moment) and via jerky changes of  $\vec\xi$ (``extrinsic'' noise).  

The
extrinsic noise is readily seen to be very small compared to that from the
person's CofM gravitational field $\Phi_{\rm CofM} = -GM \xi^{||}/r^2$:
\begin{equation}
{\sqrt{S_h^{\rm ext}} \over \sqrt{S_h^{\rm CofM}} } \sim {\Phi_{\rm ext} \over
\Phi_{\rm CofM}} \alt {9\over 2}{{\cal I}_{jk}\over Mr^2} \sim 0.01
\left({10 {\rm m}\over r}\right)^2 \;.
\label{eq:ExtrinsicNoise}
\end{equation}
Here we have used an obvious estimate of the person's quadrupole moment.

The intrinsic noise, arising from jerky changes of ${\cal I}_{jk}$, has
contributions from both terms in Eq.\ (\ref{eq:PhiQuad}).  That from the second
term is smaller by $\sim 2 \xi/r \alt 0.15 ({10 \rm m}/r)$ 
than that from the
first term, so we shall ignore it.  Taking the gradient of the first term to
obtain the gravitational acceleration on the test mass, and proceeding as in
the derivation of Eq.\ (\ref{eq:dhdt}), we obtain the following expression for
the quadrupolar noise in the interferometer as a sum over contributions from
the test masses $A=1,2,3,4$:
\begin{equation}
{d^2h \over dt^2} = \sum_A {3G\over L r_A^4} \left( {\cal I}_{jk} m_{Aj} n_{Ak}
- {5\over 2} {\cal I}_{ij} n_{Ai} n_{Aj} m_{Ak} n_{Ak} \right)\;.
\label{eq:dhdtQuad}
\end{equation}
We choose the $x$ axis along the progressive direction (direction 
$\hat\xi$ of motion), the $y$ axis along the medial (transverse horizontal)
direction, and the $z$ axis vertically upward.  Then, because $\hat m_A$ and 
$\hat n_A$ are horizontal vectors, and the body's jerky motion is in the 
$x$-$z$ plane, the noise (\ref{eq:dhdtQuad}) arises solely from two components
of the quadrupole moment,
\begin{equation}
{\cal I}_{xx} = {1\over 3}(2 I_{xx}-I_{zz})\;, \quad {\cal I}_{yy} = 
-{1\over 3} (I_{xx} + I_{zz})\;.
\label{eq:calI}
\end{equation}
Here $I_{jk}$ is the second moment of the body's mass distribution (the
integral of $\rho x'_j x'_k$ over the body).  

As the person walks, the dominant contributors to jerky changes of the
quadrupole moment are the motions of his legs.  (His arms are less massive and
jerk less.)  We divide each leg into two parts, the thigh (reaching from hip to
knee) and the shank (reaching from knee to ankle); and we approximate 
each of these as a point mass located at its center of mass, thereby making an
acceptably small error.  Measurements discussed below show that the 
quadrupolar noise at frequencies $f\sim 10$Hz arises primarily
from sudden ($\tau \sim
0.5/f \sim 50$ msec) changes of the thigh and shank 
accelerations $\Delta a_k$
each time the person's heel strikes the floor (heel down).  The corresponding
sudden change of $\ddot I_{jk}$ (second time
derivative of $I_{jk}$) is
\begin{equation}
\Delta \ddot I_{jk} = 
2 m^s {x'}^s_j \Delta a^s_k +  
2 m^t {x'}^t_j \Delta a^t_k\;,
\label{eq:dotI}
\end{equation}
where the superscripts $s$ and $t$ denote shank and thigh, $m$ is the mass of
shank or thigh, and $x'_j$ is the vector from the person's CofM to the center
of mass of shank or thigh at heel down.  

Biokinesiologists measure the motions of thigh and shank in two ways: by
videotaping markers placed on them (position measurements), and via
accelerometers placed on them (acceleration measurements); for a pedagogical
discussion see \cite{wu1}.  Because of noise introduced when taking time
derivatives of the position data, the position measurements cannot give 
reliable measures of acceleration
on the short timescales $\tau \alt 50$ms of concern to us \cite{wu1}. 
Therefore, for the thigh and shank accelerations we have relied on 
accelerometer measurements as reported by Ge Wu in Fig.\ 16-11 of Ref.\ 
\cite{wu1}.  It is obvious from that figure that the dominant contributions to
the Fourier transform $\tilde a^b_k$ of $a^b_k(t)$ (for $k=x,y,z$, $b=t,s$)
arise from sharp changes at heel down.  We have Fourier transformed $a_k^b(t)$
and found that, to within a factor $\sim 2$ over the range $2.5{\rm Hz} \alt f
\alt 25 {\rm Hz}$, $|\tilde a_k^b| \propto 1/f$.  This implies that, to
adequate accuracy for our purposes (factor 2) we can regard the quadrupolar
noise as due to sudden changes of acceleration $|\Delta a_k^b| = 2\pi f
|\tilde a_k^b|$ at heel down.

Table \ref{table:Moments} shows the values of $|\Delta a_k^b|$ at heel down
for Wu's typical 60kg individual, as inferred from our Fourier transforms of 
her Fig.\ 16-11; it also shows
the values ${x'}_k^b$ of the center of mass location of leg and shank at heel
down for a typical individual, as given in Appendix A of 
Ref.\ \cite{winter}.\footnote{Beware: 
Biokinesiologists (influenced by the
Biomechanics literature) use different axis conventions from physicists:
$y$ and $z$ are interchanged so their $y$ is vertical and $z$ is medial.}
We expect these numbers to vary,
from one adult individual to another, by no more than a factor $\sim 2$.

\begin{table}
\caption{Thigh and shank properties at heel down.  Masses $m$ (in kg)
and positions
${x',z'}$ relative to the body's center of mass (in m) were taken from 
Appendix A of \protect\cite{winter}.  
Changes $|\Delta a_j|$ of 
acceleration (in m/s$^2$), on the timescales 
$\tau \protect\alt 50$ ms, were computed from Fourier transforms of
Fig.\ 11-16 of \protect\cite{wu1}.  
}
\label{table:Moments}
\vskip15pt
\begin{tabular}{llllll}
&$m$&$x'$&$z'$&$|\Delta a_x|$&$|\Delta a_z|$\\
\tableline
Thigh&5.7&0.08&-0.3&15&$6$\\
Shank&3.5&0.2&-0.6&11&$6$\\
\end{tabular}
\end{table}

By inserting the numbers from Table \ref{table:Moments} into Eq.\
(\ref{eq:dotI}) and thence into Eq.\ (\ref{eq:calI}), and adding terms in
absolute value so as to get upper bounds, we obtain for the 
sudden (timescale $\alt 50$ ms)
heel-down changes of the second time derivative of the 
person's quadrupole moment:
\begin{equation}
|\Delta \ddot {\cal I}_{xx}| \alt 35 \hbox{kg m}^2 {\rm s}^{-2}\;, \quad
|\Delta \ddot {\cal I}_{yy}| \alt 25 \hbox{kg m}^2 {\rm s}^{-2}\;.
\label{eq:calInumbers}
\end{equation}

By then Fourier transforming Eq.\ (\ref{eq:dhdtQuad}), 
we obtain the following upper limit on the quadrupolar noise 
$|\tilde h_{\Delta a}|$ induced by the 
heel-down changes of thigh/shank acceleration: 
\begin{equation}
|\tilde h_{\Delta a}|  \alt 
{9G |\Delta \ddot{\cal I}_{xx}| \over 2(2\pi f)^5 L r^4} \;, 
\label{eq:tildehQuad}
\end{equation}
where $r$ is the distance to the nearest test mass.

During half a gait cycle (one heel down), the Fourier transform of 
the person's CofM-induced noise is [Eqs.\ (\ref{eq:d4hdt4}) and 
(\ref{eq:tildeF})]
\begin{equation}
|\tilde h_{\rm CofM}| = {2G|\Delta \dot F| \over L r^3 (2\pi f)^6}\;,
\label{eq:hcm}
\end{equation} 
where we have set the 
angle-dependent factor $|\alpha|$ to its representative value, $\sqrt2$.
Near 10 Hz, the ratio of the
quadrupolar noise (\ref{eq:tildehQuad}) to this CofM noise is
\begin{equation}
{|\tilde h_{\Delta a}| \over | \tilde h_{\rm CofM}|} \alt
{9\over4} {2\pi f\over r} {|\Delta \ddot{\cal I}_{xx}| \over | \Delta \dot
F|} \sim {1 \over 10} \left({10 {\rm m} \over r}\right)
\left({f\over 10{\rm Hz}}\right)\;.
\label{hQuadOverhcm}
\end{equation}
This is significantly less than one
independent of the factor $\sim 2$ errors and
variabilities of both noises.
Thus, as was asserted in the Introduction, the dominant noise is caused by
jerkiness in the person's CofM motion.

\section{Doors, Fists, and Vehicles}
\label{sec:AutomobileMotion}

We turn, now, to gravity gradient noise produced by the impulsive stopping
of a massive, horizontally moving object---most especially a slamming
door, a fist striking a wall, or a stopping automobile.  (The impulsive
stopping of vertical motion produces a much weaker signal than horizontal
motion; cf.\ the discussion following Eq.\ (\ref{eq:dhdt}).)  Our analysis is a
variant of that originally given by Spero \cite{spero} and reaches the same
conclusions.

Since these impulsive events are not likely to occur repetitively and
continually (by contrast with the gait cycles of human walking), they are more
appropriately analyzed as a single impulsive gravitational-wave signal than as
a stochastic noise.  The standard formula for the amplitude signal to noise
ratio $S/N$ produced by such an impulsive signal $h(t)$ in a LIGO
interferometer is (e.g., Eq. (29) of \cite{300yrs} with a
factor 2 correction)
\begin{equation}
{S^2\over N^2} = \int_0^\infty {4|\tilde h|^2\over S_h(f)} df \;,
\label{eq:SOverN}
\end{equation}
where $S_h$ is the one-sided
spectral density of the interferometer's total noise and
$\tilde h$ is the Fourier transform of the signal $h(t)$. 

The signal is given by
\begin{equation}
{d^2 h \over dt^2} = \alpha {GM\xi \over L r^3}
\label{eq:dhdt2}
\end{equation}
[Eq.\ (\ref{eq:dhdt1})], where $\alpha$ is the same angle-dependent factor as
we met for a person's CofM motion [Eq.\ (\ref{eq:alpha})], 
$M$ is the object's mass, $\xi$ is its
displacement while stopping, and $r$ is its distance 
mass from the nearest interferometer test mass.  

For a slamming door or a fist striking a wall, it is the velocity $v=\dot\xi$
that changes suddenly, by some amount $\Delta v$.  Correspondingly, the Fourier
transform of the signal $h(t)$ is 
\begin{equation}
| \tilde h | = {G M |\alpha \Delta v| \over L r^3 (2 \pi f)^4}\;.
\label{eq:tildehDoor}
\end{equation}

For the benchmark ``advanced'' LIGO interferometer, we can approximate the
noise curve (Fig.\ \ref{fig:SpectrumCM}) by
$S_h = S_o [(f_o/f)^4 + (f_o/f)^{20}]$, where $S_o = 10^{-45}/{\rm Hz}$
and $f_o = 10 {\rm Hz}$.
Inserting this $S_h$ and expression (\ref{eq:tildehDoor}) into
Eq.\ (\ref{eq:SOverN}) and integrating, we obtain
\begin{eqnarray}
{S\over N} &&\simeq
{1.2 G M |\alpha\Delta v| \over 2\pi L r^3 (2\pi f_o)^3 \sqrt{S_o
f_o}} \nonumber\\
&&\simeq 1 \left( {M \Delta v\over 5\hbox{kg m/s}}\right)
\left( {10 {\rm m} \over r} \right)^3 \;.
\end{eqnarray}
Here we have used our representative value $\sqrt2$ for $|\alpha|$. 
This is the noise level discussed in the introduction.

Turn to automobiles and other vehicles.
Under normal (non-collisional) motion, the velocity of a vehicle cannot
change significantly on a timescale of 50 ms.  The acceleration, however, {\it
can} so change, and will change most strongly when the vehicle comes to a stop,
e.g.\ when parking.  

The sudden change
$\Delta a = \Delta \ddot \xi$ of acceleration, when the vehicle comes to rest,
will produce the following Fourier transform of $h(t)$:
\begin{equation}
| \tilde h | = {G M |\alpha \Delta a| \over L r^3 (2 \pi f)^5}\;.
\label{eq:tildehVehicle}
\end{equation}
Inserting this 
$|\tilde h|$ and the advanced-interferometer $S_h(f)$ into
Eq.\ (\ref{eq:SOverN}) and integrating, we obtain
\begin{eqnarray}
{S\over N} &&\simeq 
{G M |\alpha\Delta a| \over 2\pi L r^3 (2\pi f_o)^4 \sqrt{S_o
f_o}} \nonumber\\
&&\simeq 1 \left( {M\over 2 \hbox{tonne}}\right) 
\left({|\Delta a|\over 0.6 g}\right) 
\left( {30 {\rm m} \over r} \right)^3 \;.
\end{eqnarray}
Here we have used our representative value $\sqrt2$ for $|\alpha|$, and
$g$ is the acceleration of gravity.  
This is the noise level discussed in the Introduction. 

The gravitational signal from a slamming door, striking fist, or stopping 
vehicle will be mitigated to some modest extent by an opposite signal produced
by the momentum deposited in the ``reaction mass'' (the wall, floor, and/or 
ground).  However, as for human walking, the deposited
momentum spreads over such a large spatial region in a time $1/2f \sim 50$ms, 
that the mitigation will not be significant; cf.\ Sec.\ \ref{sec:FloorMotion}
and the Appendix.

\section{Conclusions}
\label{sec:Conclusions}

In this paper we have identified what we believe to be the dominant gravity
gradient noise due to normal human activities; we have estimated its spectrum
and its strength to within accuracies of a factor $\sim 3$; and we have
discussed the implications of this noise for the size of the human exclusion
zones around LIGO's test masses in the era, ca.\ 2010, of ``advanced''
interferometers.  Until that era, human gravity gradient noise is not likely to
be a serious issue for LIGO.

Our formulas and estimates can provide a basis for the design of the facilities
of other earth-based gravitational-wave detectors.

\section*{Acknowledgments}

We are grateful to Robert Spero for triggering our interest in this problem and
for helpful discussions early in this work.
We thank Ernest L. Bontrager (Associate Director of
Engineering Research at the Pathokinesiology Service of Rancho Los
Amigos Medical Center, Downey, California) for providing us with the
force-plate data that underlie Fig.\  \ref{fig:SpectrumCM} and the analysis
in Sec.\  \ref{sec:CofMMotion}, and Ge Wu (Assistant Professor of Physical
Therapy, University of Vermont) for helpful discussions of her accelerometer
data that underlie Table \ref{table:Moments}.  We thank 
James Ipser and Lee Lindblom for information about automobile motion, and
Albert Lazzarini for providing
us with details of the LIGO site design. 
This research was supported in part by NSF Grant
PHY-9424337.

\appendix

\section{Green's Function for Spreading Momentum in Floor and Ground}

When a time varying force $F_j(t)$ is applied to the surface of the earth
at a location $\vec x'$, the time varying displacement that it produces 
in the ground at a location $\vec x$ is given by 
\begin{equation}
\zeta_i(\vec x,t) = {\partial\over\partial t} \int_{-\infty}^{+\infty} 
g^H_{ij}(\vec x,t-t';\vec x') F_j(t')dt'\;. 
\label{eq:Displacement}
\end{equation}
Here $g^H_{ij}$ is the elastodynamic Green's function for a unit-step-function
(Heaviside-function) force, $\vec F(t') = H(t') \vec e_j$.
Seismologists focus on
$g^H_{ij}$ rather than on the physicists' usual delta-function-sourced
Green's function $g_{ij} \equiv\partial g^H_{ij}/\partial t$ because 
$g^H_{ij}$'s  Heaviside steps at the various seismic
propagation fronts are more easily visualized and compared with each other than
$g_{ij}$'s delta-function spikes.  

The Green's function $g^H_{ij}$ for a homogeneous medium (``homogeneous half
space'') has been computed analytically by
Lane Johnson \cite{johnson}, up to a complicated integral; 
and Johnson has evaluated it numerically for several representative 
geometries; see
his Figs.\ 2--4.  Chao \cite{chao} has derived expressions for $g^H_{ij}$ 
when both force point and field point, 
$\vec x'$ and $\vec x$, are at the surface of a homogeneous half space; 
and Ma and Huang \cite{ma} have computed $g^H_{ij}$ for layered media.

Regardless of the nature of the medium, momentum conservation
requires that 
\begin{eqnarray} 
F_j(t) &=& 
{d\over dt} \int_{\cal V} \rho {\partial \zeta_j\over\partial t}d^3x
\nonumber\\
&=& \int_{-\infty}^{+\infty} 
{d^3\over dt^3} 
\left[ \int_{\cal V} \rho
g^H_{jk}(\vec x, t-t', \vec x')d^3x\right] F_k(t') dt'\;, \nonumber\\
\end{eqnarray}
where $\cal V$ is the entire volume of the medium.
Since this must be true for every applied force, it must be that
\begin{equation}
{d^3\over dt^3} \int_{\cal V} \rho g^H_{jk}(\vec x, t-t'; \vec x') d^3x
= \delta(t-t') \delta_{jk}\;.
\label{eq:IntgH}
\end{equation}
Causality requires
that $g^H_{ij}$ vanish everywhere for $t<t'$ and be nonzero for $\vec x$
arbitrarily near $\vec x'$ when $t-t'$ is arbitrarily small but positive;
these facts, combined with
Eq.\ (\ref{eq:IntgH}) imply
\begin{equation}
\int_{\cal V} \rho g^H_{ij}(\vec x, t-t'; \vec x') d^3x 
= {(t-t')^2\over 2} H(t-t')
\delta_{ij}\;.
\end{equation}

Now consider the noise $h(t)$ produced in a gravitational-wave interferometer
by a walking person, whose feet at location $\vec x'$ produce a
horizontal force $F_j(t)$ on the floor and thence on the ground beneath the
floor.  By (i) taking three time derivatives of Eq.\ (\ref{eq:d3h3Floor}), 
with the foor and ground 
displacement $\vec \zeta$ expressed as an integral over the elastodynamic 
Green's function [Eq.\ (\ref{eq:Displacement})], 
(ii) using force balance $d(M\dot \xi_j)/dt = -F_j$ for the
floor and person, and (iii) 
setting $d^2 F_i(t) /dt^2 = \Delta \dot
F_i \delta(t)$, where $t=0$ is a time of sharp change of jerk at the
beginning or end of the walking person's weight transfer (cf.\
Sec.\ \ref{sec:ForcePlateMeasurements}), we obtain the following:
\begin{eqnarray}
&&{d^6 h \over dt^6} = {G\over L} \sum_A m_{Ai} \Delta \dot F_k 
\left\{ \delta(t) \delta_{jk} \left(1\over
|\vec x' - \vec x_A|\right )_{,i'j'}
\right.
\nonumber\\
&&\left.
-{d^3\over dt^3}
\left[ \int_{\cal V} \rho
g^H_{jk}(\vec x,t;\vec x') \left({1\over|\vec x - \vec
x_A|}\right)_{,ij}d^3x \right] \right\}\;.
\nonumber\\
\label{eq:d6h6All}
\end{eqnarray}

This equation exhibits 
the momentum-flow features discussed in the text following Eq.\ 
(\ref{eq:d3h3Floor}): The third time derivative of the Green's function
$g^H_{jk}(\vec x,t;\vec x')$ is significantly nonzero only in the expanding
shell discussed in the text---a shell whose sharp
outer edge travels at speed $c_P$
and fuzzy inner edge a bit slower than $c_S$; cf.\ Figs 2--4 of Johnson
\cite{johnson}.  Correspondingly, the contribution of the floor and earth to
$h$ is confined to that shell.  When that shell is small compared to the
separation $|\vec x'-\vec x_A|$ between the person and the
nearest test mass $A$, the double gradients in Eq.\ (\ref{eq:d6h6All}) are
nearly equal, and momentum conservation as embodied in Eq.\ (\ref{eq:IntgH})
guarantees that
the two terms in (\ref{eq:d6h6All}) (the person noise and the floor \& 
ground noise)
will nearly cancel.  When the shell is
comparable in size to the separation $|\vec x' - \vec x_A|$, the two terms will
cancel partially but not strongly.  When the shell is large compared to $|\vec
x' - \vec x_A|$, the second term (floor \& ground noise) will be negligible
compared to the first (person noise).

Equation (\ref{eq:d6h6All}), together with the explicit expressions for
the Green's functions in Refs.\ \cite{johnson,chao,ma}, could be used
to compute quantitatively the partial cancellation of person noise
and floor \& ground noise.  We have not done so, since uncertainties
elsewhere in our modeling are comparable to or larger than the errors in the
above rough estimates.

\end{document}